\documentstyle[supertab,psfig]{l-aa}

\begin{document}

\thesaurus{(08.01.1 - 08.01.3 - 08.16.4)}

\title{Is UU Herculis a post-AGB star?}

\author{Klochkova V.G., Panchuk V.E., Chentsov E.L.}

\offprints{V.G. Klochkova}

\institute{Special Astrophysical Observatory,
           N.Arkhyz, 357147 Russia}

\date{Received 19 January 1996 / Accepted 22 July 1996}

\maketitle

\markboth{Klochkova et al. Is UU\,Her a post-AGB star?} {}

\begin{abstract}

In order to understand the evolutionary status of the anomalous supergiant
UU\,Her, the prototype of the class of variable supergiants located at high
galactic latitudes, we obtained several high-resolution spectra of this
star, with the 6m telescope, over 5 years.  This material was used for a
search of possible temporal variations of the radial velocity at the
different depths in the photosphere and for studying the chemical
composition.  The average radial velocity ${{\rm V_r \approx 130\,km/s}}$
suggests that UU\,Her belongs to the old population of the Galaxy.
No systematic dependence of the velocity on depth of the line formation
layer or on ionization and excitation potential is observed.  The radial
velocity of the ${{\rm H\alpha}}$ absorption differs strongly from the
average photospheric velocity.

The iron abundance in the photosphere of UU\,Her  is significantly lower than
that of the Sun: ${{\rm [Fe/H] = -1.32}}$.  The enhancement of nitrogen
relatively to iron content ${{\rm [N/Fe]_\odot = 0.40}}$ in combination with the
carbon underabundance ${{\rm [C/Fe]_\odot = -0.30}}$ suggests that only a
first dredge-up episode occurred.  The Na content is normal relatively to
iron, therefore there is no evidence for dredging-up of Ne-Na cycle products.
The heavy s-process metals Y, Ba are depleted relative to H and Fe, which
again implies that the third dredge-up did not occur.

From the high luminosity (${{\rm log\,g \approx 1}}$), the large radial
velocity and the chemical abundance pattern, we conclude that UU\,Her is a
low-mass halo star, but not a post-AGB star.

\end{abstract}

\section{Introduction}

In this paper, we study the characteristics of the peculiar supergiant
UU\,Her, which was proposed as the prototype of a new class of variable stars
by Sasselov (1984).  Despite several photometric and spectral investigations
carried out during last decade, the nature of the UU\,Her type stars is not
understood yet.  We first summarize briefly the main characteristics of
these stars.

The UU\,Her-type stars are supergiants (MK-classes ${{\rm Ia^+}}$, Ia, Ib)
that are located at high galactic latitudes.  Besides their location in the
Galaxy, which is peculiar for supergiants, these stars have other properties
which distinguish these objects from the normal massive supergiants of the
galactic disc.  In particular, they display both spectroscopic and
photometric variability of small amplitude and with long periods, that
undergo specific changes.  They have large spatial velocities, typical for
population~II, and, as a rule, an infrared excess caused by a circumstellar
dust envelope.  If the UU\,Her stars are truly massive stars, their existence
may point to recent star formation at high latitudes in the Galaxy.

There is observational evidence for high-velocity clouds at high latitudes
with masses and density which may be sufficient for massive star formation
(see, for example, the review by Van Woerden (1993)).  We therefore cannot
apriori reject the hypothesis that UU~Her is a massive supergiant.  Moreover,
the double-mode photometric behaviour that is observed for this object is in
an agreement with the high-mass hypothesis (Zsoldos \& Sasselov 1992).

On the other hand, low-mass stars in advanced stages of evolution (the AGB
and post-AGB stages) also attain very high luminosities that are provided by
nuclear burning in the hydrogen and helium layers.  Such a low-mass
hypothesis for the high-latitude supergiants was proposed by (Luck \& Bond
1984, Bond et al 1984) and is supported by the low metallicity of several
studied UU\,Her-type stars and related objects (Bond \& Luck 1987,
Klochkova \& Panchuk 1988a, 1988b, 1989, 1993; Luck et al, 1990;
Waelkens et al, 1991, 1992; Klochkova 1995).

A useful test of these hypotheses is provided by the determination of the
abundances of the chemical elements that take a part in nuclear processes
in the course of stellar evolution.  In the case of massive supergiants, we
may expect a metallicity near the solar one, but changes in the abundances
of the CNO and some ${{\rm\alpha - process}}$ elements, such as a Na excess.
The latter has been observed for several massive yellow supergiants in the Galaxy
and was explained  (Denissenkov \& Ivanov 1987, Denissenkov 1988) as a product
of a Ne-Na cycle. In the case of low mass stars in the AGB and post-AGB stages,
the main peculiarities of the chemical composition would be low metallicity,
changes of the CNO elements, and an excess of heavy chemical elements that owe
their origin to the s-process.

For understanding the nature of the UU\,Her type stars, it may also be
important to obtain information about radial velocities and their behavior
during a long time interval and at the different layers of the stellar
atmosphere.  Such data could be helpful to define more reliably the type of
variability and to understand the mechanism of variability.

The prototype UU\,Her has been classified F5Ib in the MK system (Lopez-Gruz \&
Garrison 1993)
and has the galactic latitude ${{\rm b=41^\circ}}$.  In contrast to other
high-latitude supergiants, UU\,Her does not display the infrared excess which
is typical for stars in a post-AGB stage of evolution (Trams et al. 1991).
Spectrophotometric (Cardelli 1989) and spectroscopic (Klochkova \& Panchuk
1989) observations of the UU\,Her show that the metallicity of this star is
significantly subsolar and the log\,g value is the same as for yellow supergiants.
Formally using  calibration ${\rm M_v - log\,g}$ which has been derived by
Klochkova \& Panchuk (1988b) for massive high luminosity stars, leads us
to the value ${{\rm M_v = -8^m}}$ for UU\,Her. Such a luminosity is
too high for a post-AGB star and contradicts to the low intensity
of the well known luminosity indicator, which is the triplet ${\rm
OI\,7773\,\AA}$ (see Fig.1).
A low value  log\,g in combination with a low metallicity lead us to suggest
a post-AGB evolutionary stage.
However, no firm conclusion could be drawn, because the expected
excesses of s-process elements were not found.

In this paper, we present the results of new spectroscopic study of the
luminosity, chemical composition and radial velocities of UU\,Her.  We were
able to obtain CCD spectra in the spectral range ${{\rm\lambda 5000 - 7000\,\AA}}$
with sufficiently high resolution for this purpose, using the echelle
spectrometer LYNX of the 6-m telescope (Panchuk et al. 1993).  These
observations allow us in particular to study the CNO abundances.

\section{Observations}

Five CCD-spectra were obtained with the echelle-spectrometer LYNX attached
at the Nasmyth-focus of the 6-m telescope and equipped with a 530x580 CCD
(Panchuk et al. 1993).  The spectral resolution is ${{\rm \delta \lambda =
0.26\,\AA}}$ in the red wavelengths region.  The main characteristics of the
spectra are summarized in Table\,1.  A Th-Ar lamp was used for the wavelength
calibration; the precision is better than 0.01\,\AA.  The hot fast rotating
star HR\,4687 was observed in order to identify telluric lines and to
separate these lines from the stellar ones.   The echelle CCD images were
corrected for bias and background.  For these procedures and for the
transformation of an echelle image into a one-dimensional spectrum, we used
the ESO-MIDAS system. The equivalent widths for most spectral lines were
measured by fitting a Gaussian to the observed using profile DECH-package by
Galazutdinov (1992).

In order to increase the time base of our observations, we also reconsidered
the two photographic spectra of the UU\,Her, which were obtained  by
Klochkova and Panchuk with the Main Stellar Spectrograph of the 6-m telescope
and were used earlier (Klochkova, Panchuk 1989) for the calculation of the
abundances of several chemical elements in the atmosphere of UU\,Her.
The spectral resolution of these photograpic spectra amounts to
${{\rm \delta\lambda = 0.35\,\AA}}$.

\begin{table}
\caption{Information on the used spectra of UU\,Her}
\begin{tabular}{cccrl}
\hline
 Date  & JD 244...   & ${{\rm \Delta\lambda,\,\AA}}$ & S/N & Detector   \\
\hline
 03.05.88.  &  7285  &   5200-6600  &  50 & 103aF   \\
 06.08.88.  &  7380  &   5200-6600  &  50 & 103aF   \\
 18.08.92.  &  8853  &   5000-7200  & 130 & CCD${{\rm^a}}$  \\
 21.08.92.  &  8856  &   5000-7200  & 120 & CCD   \\
 13.03.93.  &  9060  &   5500-8800  & 110 & CCD${{\rm^a}}$  \\
 11.05.93.  &  9119  &   5500-8800  & 110 & CCD${{\rm^a}}$  \\
 25.06.93.  &  9164  &   5500-8800  & 100 & CCD   \\
\hline
\end{tabular}
\begin{list}{}{}
\item[{\rm ''a''}] marks the spectrum which has been used for chemical
  composition determination
\end{list}
\end{table}

\section{Radial-velocity measurements}

The photograpic spectrograms were measured with an oscilloscopic comparator.
For the velocity determination from the CCD spectra, an identical procedure
of matching the direct and mirrored profiles after reduction was used
(Galazutdinov 1992).  In the case of the photographic spectrograms
independent measurements were made for several spectral regions with a length
of ${{\rm 100 - 200\, \AA}}$ long; in the case of the echelle-spectra
velocities were determined for every of the 32 orders, with a length of
${{60 -80\, \rm \AA}}$.  The wavelength calibration was performed with a
Th-Ar lamp and checked with the telluric ${{\rm O_2}}$ and ${{\rm H_2O}}$
lines.  The typical residual systematic error of an individual order, for
the spectra of 18 and 21.08.92, when the largest number of stellar lines
were observed, is 0.4\,km/s.  We consider this value as the
upper limit for the systematic error on the radial velocities listed in
Table~2.  The accuracy was successfully tested with the interstellar lines
of NaI, which show a constant velocity for the main component (23\,km/s) and
for the depression in the red wing (about -12\,km/s).

\begin{table}
\begin{center}
\caption{The results of radial velocity determination.
         The values in brackets are the number of metal lines used for
         ${{\rm V_r}}$ measurement}
\begin{tabular}{c|ccrcc}
\hline
 Date   & \multicolumn{5}{c}{${{\rm V_r,\,km/s}}$}  \\
  &  metals  & ${{\rm \sigma}}$ & n & NaI & ${{\rm H\alpha}}$   \\
\hline
 03.05.88. &  -138.5 & ${\pm 0.9}$ &  (19) &  -138.1 &  -129   \\
 06.08.88. &  -139.1 & ${\pm 0.8}$ &  (22) &  -142.0 &  -129   \\
 18.08.92. &  -125.7 & ${\pm 0.4}$ & (110) &  -126.5 &  -134  \\
 21.08.92. &  -124.7 & ${\pm 0.5}$ & (101) &  -126.0 &  -135   \\
 13.03.93. &  -139.6 & ${\pm 0.9}$ &  (39) &  -140.0 &  -136  \\
 11.05.93. &  -131.2 & ${\pm 0.7}$ &  (50) &  -131.0 &  -134  \\
 25.06.93. &  -133.3 & ${\pm 0.9}$ &  (51) &  -131.0 &  -135  \\
\hline
\end{tabular}
\end{center}
\end{table}

The second column of Table\,2 contains the mean radial velocities for the
lines Fe I,II, Ti I,II, Cr I,II, which are predominant in the UU\,Her spectra,
as well as for MgI, CaI, SiII, ScII, BaII and some other ions.  The random
errors reflect the inaccuracy of the measurements and reduction, adopted
laboratory wavelengths and some small differential shifts of some lines.
The laboratory wavelengths are taken from the solar spectrum tables
(Pierce \& Breckinridge 1974) with small corrections for the gravitational
redshift, taking into account the lower gravity of UU\,Her.

\begin{figure*}
\par
\centerline{\psfig{figure=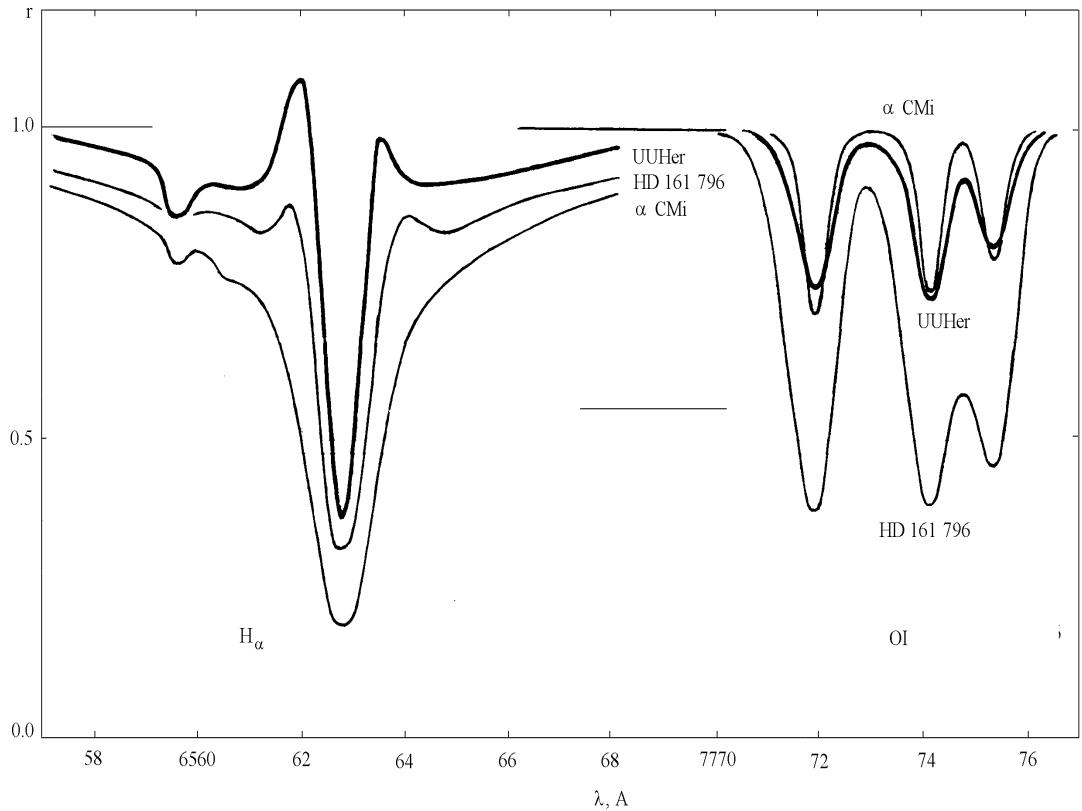,width=15.0cm,height=12.0cm}}
\par
\caption{${{\rm H\alpha}}$ and OI - profiles in the spectra
    of UU\,Her, HD\,161796 (F3Ib) and ${{\rm \alpha\,CMi}}$ (F5IV-V).}
\end{figure*}

\section{Chemical composition}

The CCD-spectra marked by "a" in Table\,1 were used for the determination
of the chemical composition of UU\,Her.  We interpolated between Kurucz`s
(1979) models, i.e. assumed local thermodynamic equilibrium (LTE) and
hydrostatic equilibrium. It is known, however, that the atmospheric
parameters of a supergiant may be altered by non-LTE effects (by
overionization mainly).  But the influence of these effects on the relative
values [X/Fe] is expected to be low, since Hill et al. (1994) estimated
abundances with and without overionization and concluded that the
differences are below 0.06\,dex.

The gf-values for most spectral lines were taken from Thevenin (1989, 1990);
for the lines of the neutral atoms (Ti, Cr, Mn, Fe) we prefer the accurate
experimental data of the Oxford group (Blackwell  et al. 1986); the
CNO-abundances were determined by using the gf-data from the Waelkens
et al. (1991).

It is obvious that the application of the standard model atmosphere method
to the analysis of spectra of a star with an extended and unstable atmosphere
can be questioned.  We therefore used only weak lines, which are formed
deeply, in our analysis.  Most of the lines we used have equivalent widths W
less than 150\,m\AA, but several ionized iron and barium lines have W near
200\,m\AA.  To control the validity of the method we analyzed in a similar
way the spectrum of the normal supergiant ${{\rm \alpha\,Per}}$, which is a
member of a young open cluster and has a solar chemical composition aside
from the non-solar [CNO/Fe] which may result from mixing of the products of
hydrogen burning.

The main difficulty in the chemical composition calculation of UU\,Her is the
determination of the effective temperature ${{\rm  T_{eff}}}$.  First, the
hydrogen line profiles in the spectra of this unstable supergiant are
distorted by dynamic processes in the extended atmosphere (Fig.\,1).  Second,
the observed irregular variability of UU\,Her restricts the application of
photometric indices for the ${{\rm T_{eff}}}$ determination too.  Therefore
we applied spectroscopic criteria for the determination of the photospheric
parameters, following an iterative procedure.  In several consecutive
iterations we determined ${{\rm T_{eff}}}$ - by forcing the independence of
the neutral iron abundance upon the line excitation potential, the surface
gravity log\,g  --  by forcing the ionization equilibrium for FeI and FeII,
and the microturbulent velocity ${{\rm \xi_t}}$ - by forcing the independence
of the abundance derived from individual FeI lines upon the equivalent width.
The errors of the model parameters are typically ${{\rm \Delta T_{eff} =
\pm 200\,K}}$, ${{\rm \Delta log\,g = \pm 0.5}}$ and
${{\rm \Delta \xi_t = \pm 0.5\,km/s}}$.

The adopted atmospheric parameters for different observing dates are listed
in Table\,3. The calculated abundances of chemical elements for three
individual spectra of UU\,Her are also given in Table\,3, together with the solar
chemical composition (Grevesse, Noels 1993). The accuracy of the abundances
depends on the accuracy of the observational data and on that of the model
fitting; the standard deviations are listed in Table\,3 for all chemical elements
studied. For the elements with a number of lines ${{\rm n > 5}}$ the
internal error $\sigma $ does not exceed 0.2.

\begin{table*}
\begin{center}
\caption{Chemical composition of the UU\,Her for  several  date  in
                  comparison with the solar one. The numbers of
                  used lines for each chemical element are given in
                  the  brackets}
\begin{tabular}{lrcr|ccr|ccr|c}
\hline
       &\multicolumn{3}{c|}{18.08.92.}&\multicolumn{3}{|c|}{13.03.93.}&
                   \multicolumn{3}{|c|}{11.05.93.} & \\
&\multicolumn{3}{c|}{${\rm T_{eff}=5700, log\,g=0.5,\xi_t=5.5}$}&
  \multicolumn{3}{|c|}{${\rm T_{eff}=6000, log\,g=0.7, \xi_t=4.0}$}&
  \multicolumn{3}{|c|}{${\rm T_{eff}=6010, log\,g=1.0,\xi_t=4.5}$} & \\
\hline
Element &${{\rm\epsilon(X)}}$ & ${{\rm\pm\sigma}}$ & n & ${{\rm\epsilon(X)}}$ &
       ${{\rm\pm\sigma}}$ & n & ${{\rm\epsilon(X)}}$ & ${{\rm\pm\sigma}}$ & n &
       ${{\rm\epsilon(X)}}$ $^{\rm a}$ \\
\hline
 Li${\rm ^b}$& 1.56 &    & (1)& 1.45&    & (1)& 1.58&    & (1) & 1.16 \\
 CI  & 7.46&0.19 & (3)& 6.65&0.15& (4)& 7.44&    & (1) & 8.55 \\
 NI  &     &     &    & 7.27&0.04& (3)& 7.12&.10 & (2) & 7.97 \\
 OI${\rm ^b}$& 7.96&0.22 & (2)& 7.58&0.05& (2)& 7.97&.37 & (2) & 8.87 \\
 OI${\rm ^f}$& 8.16& ??? & (1)& 8.03&    & (1)&     &    &     &      \\
 NaI & 4.98&0.18 & (4)& 5.17&0.24& (4)& 5.15&.24 & (4) & 6.33 \\
 MgI & 6.48&0.10 & (4)& 6.89&0.14& (5)& 7.05&.43 & (2) & 7.58 \\
 MgII&     &     &    & 6.90&0.22& (2)&     &    &     &       \\
 SiI & 6.85&0.10 &(17)& 6.88&0.04&(27)& 6.85&.06 &(14) & 7.55  \\
 SiII& 6.45&     & (1)& 6.57&    & (1)& 6.41&    & (1) &       \\
 CaI & 5.02&0.08 &(22)& 5.13&0.06&(20)& 5.16&.07 &(17) & 6.36  \\
 ScII& 1.38&0.09 &(13)& 1.52&0.06& (9)& 1.60&.08 & (9) & 3.17   \\
 TiII& 3.57&0.07 & (7)& 3.86&    & (1)& 3.64&    & (1) & 5.02   \\
 VII & 2.46&0.15 & (4)& 2.54&0.34& (2)& 2.20&.14 & (3) & 4.00  \\
 CrI & 3.94&0.05 & (8)&     &    &    &     &    &     & 5.67  \\
 CrII& 3.90&0.07 & (9)&     &    &    &     &    &     &        \\
 MnI & 4.30&0.21 & (7)& 4.07&0.08& (3)& 3.84&.12 & (3) & 5.39  \\
 FeI & 6.18&0.02&(127)& 6.34&0.02&(66)& 6.31&.02 &(71) & 7.50  \\
 FeII& 6.19&0.04& (19)& 6.36&0.05&(12)& 6.31&.03 &(12) &       \\
 CuI & 3.02&0.19 & (3)& 3.28&    & (1)& 3.14&    & (1) & 4.21 \\
 YII & 0.40&0.19 & (7)& 1.05&0.37& (2)& 1.30&.11 & (2) & 2.24  \\
 BaII& 0.14&0.08 & (3)& 0.41&0.21& (3)& 0.49&.13 & (3) & 2.13  \\
 LaII${\rm ^b}$&-0.04&0.16 & (4)&-0.06&0.13& (2)& 0.30&.20 & (4) & 1.22  \\
 CeII${\rm ^b}$&-0.12&0.21 & (3)& 0.19&    & (1)& 0.64&.46 & (2) & 1.55  \\
 PrII${\rm ^b}$&-0.45&0.10 & (2)&     &    &    &     &    &     & 0.71 \\
 NdII&   0.27&0.12   & (6)& 0.81&0.11& (4)& 0.98&.05 & (2) & 1.50  \\
 EuII&-0.42&         & (1)&-0.46&    & (1)&-0.15&    & (1) & 0.51  \\
\hline
\end{tabular}
\begin{list}{}{}
\item[$^{\rm a}$] the solar abundances (Grevesse, Noels 1993)
\item[$^{\rm b}$] the abundaces for these elements are determined only
               lines with ${{\rm W<10\,m\AA}}$  only
\item[$^{\rm f}$] abundance is determided using forbidden line ${{\rm\lambda
            6300\,\AA}}$
\end{list}
\end{center}
\end{table*}

\section{Discussion of the results}

\subsection{Radial velocities}

The light curve of UU\,Her is similar to that of RV\,Tau variables
(Zsoldos \& Sasselov 1992).  We therefore have to pay attention to the line
profile anomalies typical for such pulsating population II stars.
Unfortunately, the photometric phases corresponding to the observing epochs
of our spectrograms are not known.  In fact, the effects of pulsations on
the spectra appear to be small: obvious emission is present only in the
${{\rm H\alpha}}$ - profiles, but the profiles of the majority of absorption
lines are quite normal.  The position of lines varies, but not the widths.
The absorption lines were not observed to become asymmetric or to split
into components.

No systematic progress of the velocity with depth of the line formation zone
or with ionization and excitation potential was observed either.  Therefore
a unique value of the radial velocity, obtained  by  averaging over all
measured absorption lines except ${{\rm H\alpha}}$ and Na~I, is presented in
column\,2 of Table\,2 for each observing date.  It should be noted that our
radial velocity values fall within the interval obtained, with a different
methodology, for UU~Her by Waelkens \& Mayor (1993).  Differential shifts
exceeding the methodic errors are seen for strongest lines only: FeII\,5018\,\AA,
MgI\,5183\,\AA, NaI\,5890,96\,\AA, SiII\,6347\,\AA, Ba\,6141\,\AA, 6497\,\AA.
The asymmetry of the lines is small: the cores of the lines show both blue and
red displacements not exceeding 1-2\,km/s.  The wings tend to be displaced
blueward by 2-3\,km/s.  For the extreme case of the resonance lines of Na\,I
this is clearly seen from the comparison of columns~2 and 5 of Table\,2 and from
Fig.\,2.  It is doubtful that this blue shift of the wings is a direct
ma\-ni\-fes\-ta\-tion of the velocity gradient in the atmosphere.  It may suggest the
presence of a weak emission, that slightly raises the red wing; on the other
hand, in the ${{\rm H\alpha}}$ emission (Fig.\,1,~2) the blue component usually
dominates.

The radial velocity obtained from the ${{\rm H\alpha}}$ absorption (column~6
of Table\,2) deviates more strongly from the average photospheric velocity.
As can be seen from Fig.\,2, this absorption affects slightly both the position
and the profile of the line.  The half-width is similar to that of the
photospheric lines, and the profile has not a Stark but a Doppler shape.
Most likely, then, the ${{\rm H\alpha}}$-absorption is formed in an extended
and relatively stable (at least from August 1992 to June 1993) circumstellar
envelope.  Our limited data set does not allow us to decide whether the
${{\rm H\alpha}}$-velocity (about -135\,km/s in the echelle spectra)
corresponds to the center-of-mass velocity of the star.

\begin{figure*}
\par
\centerline{\psfig{figure=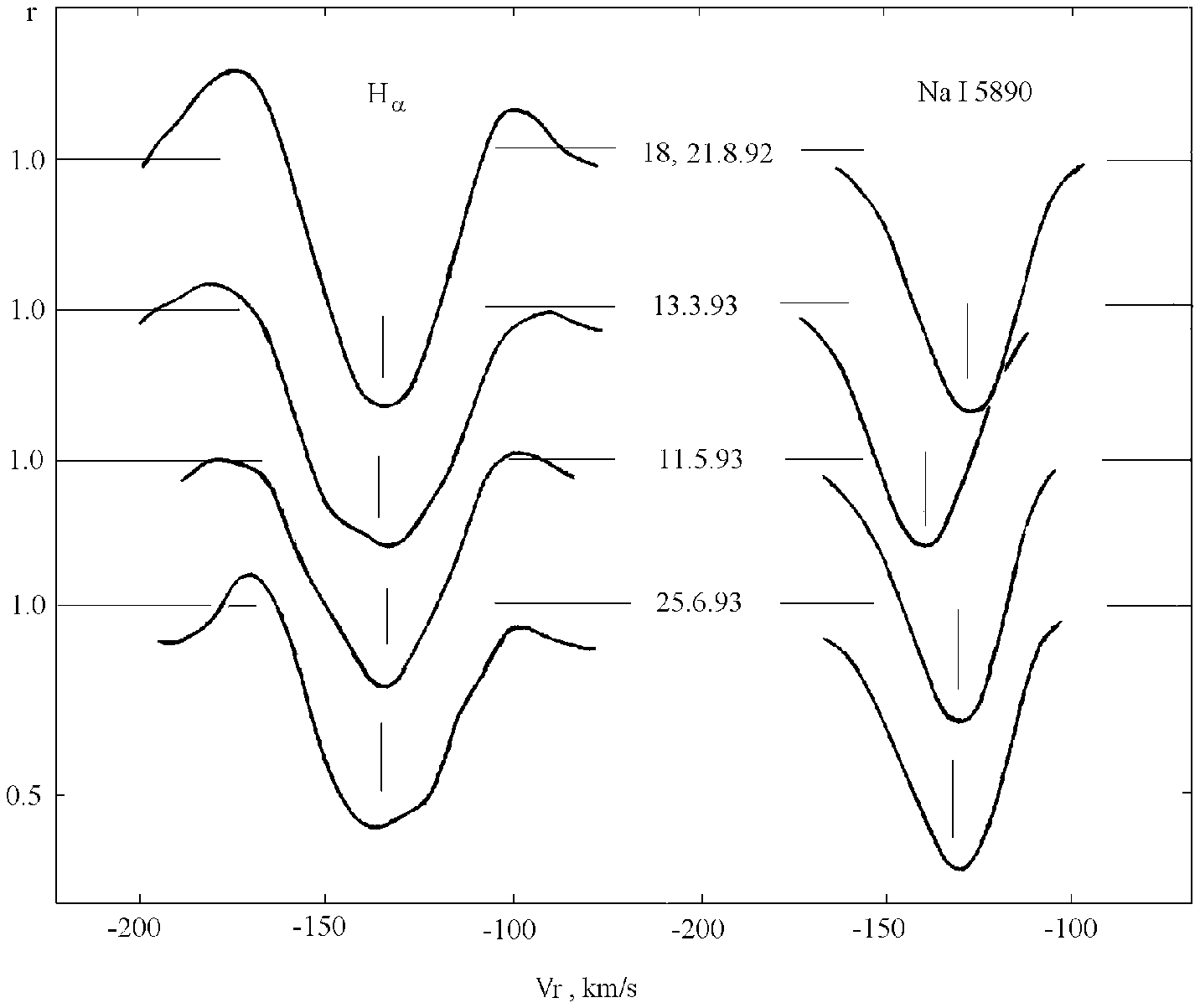,width=15.0cm,height=12.0cm}}
\par
\caption{The time variation of the profiles of the ${{\rm H\alpha}}$
               and NaI ${{\rm D_2}}$ line}
\end{figure*}

The temporal characteristics of the movements in the atmosphere of UU\,Her can
be discussed from the data for all absorption lines except ${{\rm H\alpha}}$.
Certainly, our observational data are not sufficiently numerous for an
independent estimate of a characteristic timescale.  We note that during three
days (from August 18 to August 21 1992) the velocity almost did not change,
but that over ten months it varied by some 15\,km/s.  From the photometry it is
known that pulsations occur with the 73-days fundamental mode and the 45-days
first overtone (Zsoldos \& Sasselov 1992); the spectroscopy suggests that this
oscillations are accompanied by velocity variations with some 15\,km/s and with
a cycle duration no less than 1 month.  The need for simultaneous observations
of both kinds is obvious.

\subsection{Chemical abundance pattern}

The details of the chemical abundances of UU\,Her are considered with respect
to the iron content ${{\rm\epsilon(Fe)}}$:\\

${{\rm [X/Fe] =  [log\epsilon(X) - log \epsilon (X)]_{\star} -
             [log\epsilon(X) - log \epsilon (X)]_{\sun}}}$.\\

\noindent  The values [X/Fe] for UU\,Her are presented in Table\,4.
For comparison we there also list results for the normal massive supergiant
${{\rm \alpha Per}}$ (Klochkova 1995) and for the post-AGB star ROA24 in the
the globular cluster ${{\rm \omega\,Cen}}$ (Gonzalez \& Wallerstein 1992).
Below we compare the chemical composition of these three high-luminosity
objects.  In Table\,5 we present the W-value of lines obtained from one of our
spectra of UU\,Her.  Given there are also the gf-values and the results of the
abundance calculations of the chemical elements that are most important
for our purpose.

\begin{table}
\caption{The average value [X/Fe]  for UU\,Her in comparison to that
        for the normal supergiant ${{\rm\alpha\,Per}}$ and
        post-AGB star ROA24 from  the globular cluster}
\begin{tabular}{lcrr}
\hline
 Element & \multicolumn{3}{c}{[X/Fe]}          \\
\hline
  & UU\,Her& ${{\rm\alpha\,Per^a}}$ & ROA24$^{\rm b}$ \\
\hline
 Li           &  1.69           &        &       \\
 CI           & -0.30           & -0.17  &   0.67 \\
 NI           &  0.40           &  0.65  &   1.02 \\
 OI           &  0.19           & -0.27  &   1.01 \\
 OI$^{\rm f}$ &  0.46           & -0.35  &        \\
 NaI          & -0.01           &  0.23  &   0.71 \\
 MgI          &  0.45           &  0.22  &   0.31 \\
 MgII         &  0.47           &        &   0.09 \\
 SiI          &  0.53           &  0.19  &   0.80 \\
 SiII         &  0.15           &  0.55  &   1.03 \\
 CaI          &  0.04           &  0.15  &   0.60 \\
 ScII         & -0.45           & -0.36  &  -0.13 \\
 TiII         & -0.12           &  0.00  &   0.33 \\
 VII          & -0.49           & -0.28  &   0.15 \\
 CrI          & -0.41           &  0.23  &   0.65 \\
 CrII         & -0.45           &  0.08  &  -0.01 \\
 MnI          & -0.02           & -0.06  &   0.35 \\
 FeI          &  0.00           & -0.01  &   0.00 \\
 FeII         &  0.01           &  0.02  &   0.00 \\
 CuI          &  0.16           &  0.53  &        \\
 YII          & -0.20           &  0.02  &  +0.37 \\
 BaII         & -0.56           &  0.01  &  +0.96 \\
 LaII         &  0.11           & -0.02  &  +0.54 \\
 CeII         &  0.07           &        &  +1.60 \\
 PrII         &  0.16           &        &        \\
 NdII         &  0.40           &  -0.52 &  +0.67 \\
 EuII         &  0.37           &   0.09 &  +0.25 \\
\hline
\end{tabular}
\begin{list}{}{}
\item[$^{\rm a}$] the data for ${{\rm\alpha\,Per}}$ from the paper of Klochkova
                  (1995).
\item[$^{\rm b}$] the data for ROA24 from the paper of Gonzalez and
                  Wallerstein (1992)
\item[$^{\rm f}$] abundance is determided using forbidden line ${{\rm \lambda
            6300\,\AA}}$
\end{list}
\end{table}

\begin{description}

\item[{\it Lithium.}]  The lithium doublet near ${{\rm \lambda 6707\,\AA}}$ is
very weak (on average ${{\rm W = 6\,m\AA}}$) in the UU~Her spectra, but it is
confidently measured in all three studied spectra. The calculated Li abundance
is consistent with an advanced evolution stage of a halo star.

\item[{\it CNO-group.}]  The behaviour of this triad is a major test of the
evolutionary status.  As is shown in Tables\,3 and 5, the abundances of the CNO
elements in the atmosphere of UU\,Her could be determined accurately from
several lines.  An underabundance of carbon and overabundance of nitrogen
relatively to iron are obtained.  An underabundance of carbon is also found for
the standard supergiant ${{\rm \alpha\,Per}}$, in according to an evolutionary
stage where this star went through the first dredge-up.  On the other hand,
the post-AGB star ROA24 is overabundant in both carbon and nitrogen.

The oxygen abundance is different for the three objects.  For
${{\rm \alpha\,Per}}$ oxygen is underabundant with respect to iron.  For ROA24
oxygen is much strengthened relatively to iron, due to He burning at the red
giant stage and following mixing and dredging-up.  For UU\,Her the OI abundance
is based on the very weak lines near ${{\rm \lambda 6155\,\AA}}$; it is somewhat
strengthened relatively to iron.  We thus conclude that the UU\,Her evolutionary
stage is not so advanced as for ROA24: it appears that the third dredge-up has
not occured yet.  It should be noted that the abundance of OI calculated by
using the lines of the IR-triplet near ${\rm \lambda 7773\,\AA}$ is larger
still (see Table\,5).  It is known that the lines of this triplet are formed in
non-LTE conditions in the atmosphere of high luminosity star (Faraggiana et
al. 1988, Kiselman 1993).  Therefore we did not use these lines for the
calculation of the average ${\rm \epsilon(O)}$.  It should be repeated
that the intensity of OI triplet in the UU\,Her spectrum is not so high as
could be expected for a supergiant with ${\rm log\,g \approx 1.0}$ (see
Fig.\,1).
The total equivalent width of the OI triplet lines is
${\rm \Sigma W(OI) = 0.527\,\AA}$, in agreement with the results of
Arellano Ferro \& Mendoza (1993), who discussed a sample of high-latitude
A-G supergiants and concluded that for UU\,Her the luminosity from the
${\rm OI\,7773\,\AA}$ intensity is not so high as from
${\rm uvby,\beta}$-photometry.

\item[{\it Light metals.}]
The abundance of some light metals produced by the ${{\rm \alpha}}$-process are
enhanced for the UU\,Her: ${{\rm [Mg/Fe] = 0.45}}$, ${{\rm [Si/Fe] = 0.53}}$.
At the same time Na and Ca are significantly underabundant.
The odd-even effect is present: ${{\rm [Na/Mg] = -0.46}}$.  The sodium and
calcium behaviour of UU\,Her is different from that for ROA24.  The sodium
content is especially interesting, because it is known that it is often
strengthened for supergiants with masses larger than ${{\rm 2 - 3}
{\cal M}_\odot}$, where it is an indication for the dredge-up of matter
to the surface of star (Denissenkov 1988).  The lack of a Na-excess is thus an
additional argument for the hypothesis that UU\,Her is an evolved low-mass
metal-deficient star.

\item[{\it Metallicity.}]
The iron abundance is determined for UU\,Her with a high precision from a
sample of 60-120 FeI lines and more than ten FeII lines.  The average
metallicity is equal to ${\rm [Fe/H]_\odot = -1.32}$.  The ratio for
manganese is close to the solar one: ${\rm [Mn/Fe] = -0.02}$.
The abundance of the iron group elements (Sc, Ti, V, Cr) are slightly
decreased relatively to the Fe content.  An average
${\rm[X/Fe] \approx -0.38}$  is found for these elements.   Since the
abundances of these iron-group elements are not altered during nuclear
processes we suppose that the systematic decreasing of ScII, TiII, CrII is
the manifestation of the departure from the LTE conditions.  The behavior of
these elements is similar for the standard star ${\rm \alpha\,Per}$.
Probably, as noted by Barker et al. (1971), there is any additional source
of ionization in the atmospheres of high-luminosity stars for these ions with
a low second ionization potential.
It should be noted that the higher ratio [Mn/Fe] close to the solar
one may be explained by the uncounted superfine structure of the atomic
levels.

\item[{\it Heavy metals.}]
From the s-process elements we measured accurately three BaII lines (see
Table\,5).  The abundance from these individual BaII lines are in a good
agreement.  We conclude that Ba is deficient and not in excess for UU\,Her.
Also the s-process element Y appears to be underabundant, but due to the
weakness of its lines this result is less definitive than for Ba.
For the normal supergiant ${{\rm \alpha\,Per}}$ there is no s-process
peculiarity, but for ROA24 a prominent excess of Ba and Y is observed, as is
expected for a low mass star that has undergone third dredge-up.

The abundance of lanthanides (La, Ce, Nd, Pr) and of Eu, elements that are
mainly produced by the r-process, are slightly enhanced or normal relatively
to iron.  For these heavy metals the average value is
${{\rm [X/Fe] = +0.22\pm 0.06}}$.

\end{description}

Considering all evidence, the chemical composition of UU\,Her is close to the
average for the sample of UU\,Her candidates discussed by Klochkova
\& Panchuk (1993).

\section{Conclusions}

Spectroscopically, UU\,Her is a supergiant, because of its low surface gravity
${{\rm log\,g \approx 1.0}}$.  Its radial velocities reveal the oscillations
with an amplitude of approximately 15\,km/s and a cycle duration of
approximately one month.

The main results of the chemical composition research presented in this
paper are as follows:
\begin{itemize}
    \item an overall deficiency of metals,
    \item a slight overabundance of N in combination with a C depletion,
    \item an overabundance with respect to iron of the even-Z elements
    (Mg, Si) synthetized by the $\alpha $-process,
    \item an abundance (with respect to iron) close to the solar one of the
    other $\alpha $-process elements Na and Ca,
    \item a strong odd-even effect,
    \item the absence of any traces on the third dredge-up.
\end{itemize}

Therefore, from the chemical abundance pattern, taking into account the high
radial velocity, the absence of a detached envelope and IR-excess, we conclude
that UU\,Her did not yet undergo a significant mass loss episode and convection
development.

\begin{acknowledgements}

We are much indebted to Dr.~C.\,Waelkens for its helpful and valuable remarks
at the refereeing of the manuscript.

The project ``Spectroscopic research of stars evolved from the AGB to
planetary nebulae''  has the  financial support from the Russian Federal
Program ``Astronomy''.

\end{acknowledgements}

\begin{center}
\tablecaption{Data of the measurements and calculations for the one (13.03.93)
              from the UU\,Her spectra obtained}
\tablehead{\hline $\lambda $ & gf & ${\rm W\,(m\AA)}$
                          & ${\rm\epsilon(X)}$ \\
    \hline \rule{0pt}{5pt}&&&\\}
\tabletail{\hline}
\begin{supertabular}{crrl}
\multicolumn{4}{l}{Li}\\
           6707.78  &        0.02 &    4    &-10.55  \\
\multicolumn{4}{l}{C}\\
           7113.17  &       -0.86 &   11    & -5.09  \\
           7115.19  &       -0.90 &    9    & -5.13  \\
           8335.16  &       -0.48 &   72    & -5.47  \\
           8727.13  &       -8.21 &   17    & -5.71  \\
\multicolumn{4}{l}{N}\\
           8629.16  &        0.03 &   21    & -4.67  \\
           8711.64  &       -0.14 &   24    & -4.72  \\
           8718.76  &       -0.17 &   19    & -4.81  \\
\multicolumn{4}{l}{O}\\
           6155.99   &      -0.66 &    4    & -4.47  \\
           6156.78   &      -0.44 &    7    & -4.38  \\
           6300.32   &      -9.76 &    21   & -3.97  \\
           7771.96   &       0.29 &   191   & -3.55  \\
           7774.17   &       0.14 &   204   & -3.29  \\
           7775.39   &      -0.14 &   132   & -3.73  \\
\multicolumn{4}{l}{Na}\\
           5682.63   &      -0.60 &    18   & -7.31  \\
           5688.20   &      -0.15 &    59   & -6.97  \\
           6154.22   &      -1.66 &    14   & -6.18  \\
           6160.75   &      -1.35 &     6   & -6.87  \\
\multicolumn{4}{l}{Mg}\\
           5711.09   &      -1.75 &    42   & -5.39  \\
           6318.70   &      -1.96 &     8   & -5.27  \\
           6319.24   &      -2.30 &     9   & -4.89  \\
           8310.25   &      -1.14 &     7   & -5.54  \\
           8717.83   &      -0.88 &    42   & -5.00  \\
\multicolumn{4}{l}{MgII}\\
           7877.05   &       0 40 &    64   & -4.88  \\
           7896.37   &       0.74 &    56   & -5.33  \\
\multicolumn{4}{l}{Si}\\
           5645.66   &      -2.15 &    13   & -5.24  \\
           5665.60   &      -2.11 &    15   & -5.26  \\
           5666.69   &      -1.67 &     9   & -5.44  \\
           5684.52   &      -1.66 &    63   & -4.80  \\
           5690.47   &      -1.91 &    36   & -4.97  \\
           5708.44   &      -1.49 &    50   & -5.00  \\
           5753.64   &      -1.45 &    11   & -5.16  \\
           5772.26   &      -1.78 &    39   & -4.74  \\
           5797.86   &      -2.03 &    15   & -5.06  \\
           5948.54   &      -1.22 &    71   & -4.96  \\
           6087.79   &      -1.80 &     4   & -5.04  \\
           6091.92   &      -1.33 &     7   & -5.34  \\
           6142.49   &      -1.56 &    11   & -5.17  \\
           6145.08   &      -1.48 &    18   & -5.00  \\
           6155.14   &      -0.84 &    58   & -5.00  \\
           6237.34   &      -1.22 &    33   & -4.98  \\
           6243.82   &      -1.34 &    19   & -5.15  \\
           6244.47   &      -1.38 &    15   & -5.26  \\
           6976.50   &      -1.07 &    14   & -5.16  \\
           7003.57   &      -0.86 &    20   & -5.21  \\
           7405.78   &      -0.71 &    52   & -5.20  \\
           7415.96   &      -0.67 &    50   & -5.26  \\
           7799.99   &      -0.76 &     9   & -5.48  \\
           7918.39   &      -0.58 &    60   & -4.90  \\
           8556.79   &      -0.21 &    67   & -5.44  \\
           8752.02   &      -0.37 &    50   & -5.47  \\
\multicolumn{4}{l}{SiII}\\
           6347.09   &       0.31 &   123   & -5.43  \\
\multicolumn{4}{l}{Ca}\\
           5581.97   &      -0.63 &    14   & -7.37  \\
           5590.12   &      -0.74 &    12   & -7.33  \\
           5594.47   &      -0.31 &    73   & -6.73  \\
           5598.49   &      -0.35 &    64   & -6.79  \\
           5601.28   &      -0.63 &    20   & -7.16  \\
           5857.45   &       0.07 &    65   & -6.78  \\
           6102.72   &      -0.80 &    64   & -6.91  \\
           6122.22   &      -0.20 &   123   & -6.96  \\
           6162.17   &      -0.10 &   110   & -7.16  \\
           6163.75   &      -1.46 &    15   & -6.46  \\
           6166.44   &      -1.26 &    14   & -6.71  \\
           6169.06   &      -0.75 &    16   & -7.14  \\
           6169.56   &      -0.57 &    30   & -6.99  \\
           6439.07   &      -0.05 &    98   & -6.72  \\
           6449.81   &      -0.62 &    24   & -7.02  \\
           6471.66   &      -0.88 &    53   & -6.33  \\
           6493.78   &      -0.39 &    66   & -6.69  \\
           6499.65   &      -1.00 &    27   & -6.58  \\
           7148.15   &      -0.08 &    42   & -7.11  \\
           7202.20   &      -0.34 &    78   & -6.45  \\
\multicolumn{4}{l}{ScII}\\
           5640.97   &      -1.02 &    33   &-10.68  \\
           5657.87   &      -0.55 &   101   &-10.42  \\
           5667.16   &      -1.21 &    32   &-10.53  \\
           5669.03   &      -1.10 &    44   &-10.44  \\
           5684.19   &      -1.08 &    52   &-10.37  \\
           5854.31   &      -2.23 &     8   &-10.17  \\
           6245.63   &      -1.15 &    40   &-10.44  \\
           6279.75   &      -1.28 &    14   &-10.76  \\
           6320.87   &      -1.89 &     6   &-10.53  \\
\multicolumn{4}{l}{TiII}\\
           6606.95   &      -2.85 &     21  & -8.14  \\
\multicolumn{4}{l}{VII}\\
           5819.93   &      -1.80 &     2   & -9.81  \\
           5928.88   &      -1.74 &    11   & -9.12  \\
\multicolumn{4}{l}{MN}\\
           6013.50   &      -0.25 &    14   & -7.84  \\
           6016.64   &      -0.24 &    13   & -7.87  \\
           6021.80   &       0.03 &    15   & -8.08  \\
\multicolumn{4}{l}{Fe}\\
           5569.62   &      -0.40 &   125   & -5.65  \\
           5572.85   &      -0.22 &   160   & -5.51  \\
           5576.10   &      -0.73 &    72   & -5.81  \\
           5586.76   &      -0.12 &   160   & -5.63  \\
           5615.65   &       0.00 &   163   & -5.76  \\
           5624.55   &      -0.65 &   106   & -5.60  \\
           5638.27   &      -0.85 &    33   & -5.45  \\
           5641.44   &      -1.18 &     8   & -5.83  \\
           5679.02   &      -0.85 &    11   & -5.70  \\
           5686.53   &      -0.64 &    21   & -5.63  \\
           5701.55   &      -2.22 &    31   & -5.67  \\
           5705.99   &      -0.58 &    22   & -5.61  \\
           5731.76   &      -1.19 &    12   & -5.72  \\
           5753.14   &      -0.66 &    19   & -6.01  \\
           5775.09   &      -1.23 &    15   & -5.36  \\
           5778.46   &      -3.60 &     2   & -5.42  \\
           5816.38   &      -0.69 &    16   & -5.57  \\
           5856.08   &      -1.69 &     4   & -5.44  \\
           5859.61   &      -0.60 &    25   & -5.47  \\
           5862.36   &      -0.38 &    25   & -5.69  \\
           5883.81   &      -1.37 &    17   & -5.49  \\
           5930.19   &      -0.23 &    22   & -5.82  \\
           5934.65   &      -1.26 &    21   & -5.48  \\
           5952.72   &      -1.51 &    12   & -5.41  \\
           6003.01   &      -1.12 &    38   & -5.37  \\
           6008.56   &      -0.98 &    37   & -5.52  \\
           6020.17   &      -0.21 &    39   & -5.60  \\
           6024.07   &      -0.02 &    56   & -5.63  \\
           6027.06   &      -1.15 &    11   & -5.74  \\
           6065.49   &      -1.53 &    63   & -5.86  \\
           6137.70   &      -1.40 &    95   & -5.72  \\
           6151.62   &      -3.30 &     6   & -5.64  \\
           6157.73   &      -1.28 &    17   & -5.46  \\
           6173.34   &      -2.88 &    12   & -5.77  \\
           6191.56   &      -1.48 &    76   & -5.96  \\
           6219.29   &      -2.43 &    47   & -5.55  \\
           6230.73   &      -1.28 &   105   & -5.79  \\
           6232.64   &      -1.33 &    30   & -5.54  \\
           6246.33   &      -0.70 &    58   & -5.84  \\
           6252.56   &      -1.69 &    60   & -5.95  \\
           6254.26   &      -2.38 &    44   & -5.56  \\
           6256.36   &      -2.34 &    26   & -5.72  \\
           6265.14   &      -2.55 &    39   & -5.56  \\
           6301.50   &      -0.59 &    49   & -6.01  \\
           6302.49   &      -1.16 &    50   & -5.41  \\
           6318.02   &      -2.00 &    42   & -5.83  \\
           6322.69   &      -2.43 &    21   & -5.65  \\
           6335.34   &      -2.20 &    44   & -5.85  \\
           6336.84   &      -0.68 &    38   & -6.06  \\
           6344.15   &      -2.92 &    15   & -5.52  \\
           6355.04   &      -2.38 &    13   & -5.82  \\
           6358.69   &      -4.47 &    12   & -5.43  \\
           6393.60   &      -1.57 &    60   & -6.02  \\
           6408.02   &      -1.00 &    25   & -5.89  \\
           6411.66   &      -0.49 &    63   & -5.92  \\
           6419.95   &      -0.27 &    19   & -5.77  \\
           6421.36   &      -2.03 &    62   & -5.69  \\
           6430.85   &      -2.01 &    77   & -5.66  \\
           6494.98   &      -1.27 &   105   & -5.94  \\
           6518.37   &      -2.67 &    13   & -5.32  \\
           6592.92   &      -1.60 &    75   & -5.58  \\
           6593.88   &      -2.42 &    53   & -5.26  \\
           6609.11   &      -2.69 &    21   & -5.36  \\
           6677.99   &      -1.22 &    66   & -6.08  \\
           6750.15   &      -2.62 &    28   & -5.43  \\
           6855.16   &      -0.63 &    23   & -5.54  \\
\multicolumn{4}{l}{FeII}\\
           5991.38   &      -3.76 &    78   & -5.52  \\
           6084.10   &      -3.99 &    40   & -5.67  \\
           6113.33   &      -4.26 &    30   & -5.53  \\
           6149.24   &      -2.88 &    78   & -5.72  \\
           6238.38   &      -2.87 &    70   & -5.83  \\
           6247.56   &      -2.55 &   110   & -5.77  \\
           6416.90   &      -2.86 &    54   & -5.97  \\
           6432.65   &      -3.85 &    94   & -5.54  \\
           6446.40   &      -2.11 &     8   & -5.51  \\
           6516.05   &      -3.55 &   100   & -5.80  \\
           7479.70   &      -3.77 &    26   & -5.50  \\
           7515.88   &      -3.57 &    54   & -5.29  \\
\multicolumn{4}{l}{Cu}\\
           5782.13   &      -1.81 &    14   & -8.72  \\
\multicolumn{4}{l}{YII}\\
           5728.91   &      -1.21 &     3   &-11.32  \\
           6795.43   &      -1.59 &    12   &-10.58  \\
\multicolumn{4}{l}{BaII}\\
           5853.68   &      -0.92 &    81   &-11.53  \\
           6141.72   &      -0.27 &   158   &-11.98  \\
           6496.90   &      -0.07 &   202   &-11.26  \\
\multicolumn{4}{l}{LaII}\\
           6262.25   &      -1.45 &     6   &-12.19  \\
           6526.95   &      -1.58 &     6   &-11.93  \\
\multicolumn{4}{l}{CeII}\\
           5610.24   &       0.00 &    10   &-11.81  \\
\multicolumn{4}{l}{NdII}\\
           5740.87   &      -0.43 &     7   &-11.47  \\
           5842.38   &      -0.34 &    11   &-11.24  \\
           6031.30   &      -0.42 &    13   &-11.13  \\
           6034.22   &      -0.40 &    12   &-10.93  \\
\multicolumn{4}{l}{EuII}\\
           6437.64   &       0.05 &    12   &-12.46  \\
\end{supertabular}
\end{center}

\end{document}